\documentclass[prd,aps]{revtex4-2}
\usepackage{amsmath,graphicx,bm,dcolumn,hyperref,amsfonts,bbold,xcolor}

\begin{document}


\title{Symanzik Effective Action for \\
Karsten-Wilczek Minimally Doubled Fermions}

\author{Kunal Shukre}
\email{kunalsatyajit.shukre@niser.ac.in}
\affiliation{%
 National Institute of Science Education and Research (NISER),\\
 Bhubaneswar 752050, P.O. Jatni, Khurda, Odisha, India\\
 and Homi Bhabha National Institute, India}%
\altaffiliation[Also at ]{Homi Bhabha National Institute,
Anushakti Nagar 400085, Mumbai, India}
\author{Subhasish Basak}%
\email{sbasak@niser.ac.in}
 \affiliation{%
 National Institute of Science Education and Research (NISER),\\
 Bhubaneswar 752050, P.O. Jatni, Khurda, Odisha, India\\
 and Homi Bhabha National Institute, India}%
\altaffiliation[Also at ]{Homi Bhabha National Institute,
Anushakti Nagar 400085, Mumbai, India}

\date{\today}

\begin{abstract}
Karsten-Wilczek (KW) fermions are a popular variant of minimally
doubled fermions. We construct Symanzik effective action for KW
fermions, which is known to break the hypercubic symmetry of the
lattice action. In this work we make the two fermionic modes,
called tastes, explicit using the point-splitting proposal of Creutz
and Misumi and write the KW action in terms of the taste fields We
identify the symmetries of the point-split action and write down
the Symanzik effective action up to dimension-5, including the
divergent dimension-3, operators for both free and interacting KW
fermions.  
\end{abstract}

\maketitle

\section{\label{sec:Intro}Introduction}
Chiral perturbation theory ($\chi$PT) \cite{Gasser:1983yg} is extensively used to explore
continuum QCD at low energies and is an essential component in
extracting physical observables from lattice simulations. On a
discrete spacetime lattice, the first step is construction of
Symanzik effective theory \cite{symanzik1983continuum,symanzik1983continuum2}
describing the lattice actions near the continuum limit using higher
dimensional operators in orders of lattice spacing $a$ which appears
explicitly in the action. Such a chiral Lagrangian is an expansion
in the quark masses (like continuum $\chi$PT) and in the lattice spacing.
This formalism is known as lattice $\chi$PT. Well-known examples of
lattice $\chi$PT are Wilson $\chi$PT \cite{bar2004chiral,rupak2002chiral,
sharpe1998spontaneous} and Staggered $\chi$PT \cite{lee1999partial}.
In this work, we build the Symanzik effective theory for Minimally
doubled fermions as proposed by Karsten and Wilczek
\cite{Karsten:1981gd,PhysRevLett.59.2397}.

According to the Nielsen-Ninomiya No-Go Theorem \cite{nielsen1981no},
fermion actions that have an exact chiral symmetry must have at least
two doublers. The doublers are multiple fermionic species on lattice
arising due to discretising spacetime. To have the chiral fermions
on the lattice, Domain-wall proposal \cite{Kaplan:1992bt} introduces
a fifth dimension while the Ginsparg-Wilson proposal
\cite{Ginsparg:1981bj} suggests redefinition of chiral symmetry on
the lattice as a go-around. Another class of lattice fermions that
are explored recently are Minimally doubled fermions.  Karsten-Wilczek
(KW) fermions are a variant of minimally doubled
fermions \cite{Creutz:2010cz}, which breaks the hypercubic symmetry of the lattice action.
Another popular variant of minimally doubled fermions is Borici-Creutz
fermions \cite{Borici:2007kz,borici2008minimallydoubledfermionrevival,
Creutz:2007af}. Numerical investigations have been carried out for different variants of MDF by various groups \cite{Kishore:2025fxt,Durr:2020yqa,Weber:2013tfa,Basak:2017oup} .  It is therefore imperative to explore the chiral structure of MDF. In this paper, we focus our study on KW fermions.

In minimally doubled fermions, the fermionic field $\psi$ describes
the two doubler fermions, which we call \textit{tastes}. The two
tastes residing at separate poles of the KW propagator can be made
explicit by \textit{point-splitting} of the field $\psi$. This
procedure of point-splitting is not unique, various suggestions for
point-splitting exist and can be found in \cite{Creutz:2010qm,
Creutz:2010bm}, \cite{Tiburzi:2010bm}, and very recently
\cite{Weber:2025kcl}. In \cite{Shukre:2024tkw}, we investigated
Tiburzi's point-splitting for KW fermions and attempted to write
the Symanzik effective action and the chiral Lagrangian. However, physical and computational problems emerge from the fact that Tiburzi's tastes are infinitely non-local. In this
paper, we used point-splitting suggested by Creutz-Misumi \cite{Creutz:2010qm,
Creutz:2010bm}.

One distinct advantage of Creutz point-splitting is that in the
configuration space the relation between the taste fields, later
denoted by $(u,\,d)$ and originally defined in the momentum space,
and the field $\psi$ is fairly intuitive. Furthermore, one can create
a taste isospinor from the taste fields and use it to define the
point-split action. Besides, the taste fields thus defined are local
and hence can be useful for numerical simulations. In Creutz's point-splitting, the mass term for the two tastes is not diagonal in the taste space and, therefore, in the continuum limit the tastes do not have the usual physical flavor states interpretation. We write the
lattice action in terms of the isospinor field with the tastes as
its components and identify the symmetries of this action. These
symmetries are used to construct the actions for the Symanzik
effective theory. Symanzik effective theory is an expansion in the
lattice spacing and hence is crucial in performing the continuum
extrapolation. In this work, Symanzik effective theory has been
constructed up to dimension-5 operators for Creutz point-split KW
fermions. Free and interacting cases have been treated separately
and Symanzik effective theories have been constructed for both.

This paper has been organized as follows: in
Sec.~\ref{sec:creutz_point_splitting}, we describe the KW action
followed by a review of Creutz's proposal for point-splitting. We
use the taste fields for constructing the point-split action and
identify its symmetries. In Sec.~\ref{sec:free_symanzik_action} we
construct the Symanzik effective action for the free KW fermions.
Following, in Sec.~\ref{sec:interacting_symanzik_action}, we study
the transformations of the link variables and construct the Symanzik
effective action for interacting KW fermions. We conclude the
discussion of our work with a summary. In Appendix A, we present
some calculations of symmetry transformations of some of the
operators as example. In Appendix B, we present a brief discussion
on Tiburzi's point-splitting.

\section{\label{sec:creutz_point_splitting}Creutz Point-Splitting Relations}
The free Karsten-Wilczek action on lattice \cite{Creutz:2010qm},
where $\hat{e}_4$ is the unit vector in the time direction, is given
in Eq.~(\ref{eq:creutz_point-split_free_action}),
\begin{equation}
\begin{split}
    \mathcal{S}=\frac{1}{2}\sum\limits^3_{j=1}\sum\limits_x\bigg[
\overline{\psi}_x\gamma_j\psi_{x+\hat{j}}-\overline{\psi}_{x+\hat{j}}
\gamma_j\psi_{x}\bigg]-\frac{i}{2}\sum\limits_x\bigg[\overline{\psi}_x
\gamma_4\psi_{x+\hat{e}_4}+\overline{\psi}_{x+\hat{e}_4}\gamma_4\psi_{x}
\bigg]\\
    +m_0\sum\limits_x\overline{\psi}_x\psi_x-\frac{i}{2}\sum
\limits^3_{j=1}\sum\limits_x\bigg[\overline{\psi}_x\gamma_4\psi_{x+
\hat{j}}+\overline{\psi}_{x+\hat{j}}\gamma_4\psi_{x}-2\overline{\psi}_x
\gamma_4\psi_x\bigg]\label{eq:free_lattice_action}
\end{split}
\end{equation}
\noindent
The corresponding Dirac operator in the momentum space is,
\begin{equation}
    D(p)=i\sum^{3}_{j=1}\gamma_j\sin(p_j)+i\gamma_4\bigg(3-
\sum^4_{\mu=1}\cos {p_\mu}\bigg) \label{eq:kwmom_action}
\end{equation}
and the poles of the propagator $D^{- 1}(p)$ are at $\Vec{p}=0$,
$p_4=\pm\pi/2$. The doublers or tastes residing at the poles are
separated using Creutz point-splitting relations, 
\begin{eqnarray}
 u(q) &=& \frac{1}{2}\bigg(1+\sin (q_4+\frac{\pi}{2})\bigg)\psi(q+
\frac{\pi}{2}\hat{e}_4),\\
 d(q) &=& \frac{1}{2}\Gamma\bigg(1-\sin (q_4-\frac{\pi}{2})\bigg)
\psi(q-\frac{\pi}{2}\hat{e}_4).\label{eq:creutz_point_splitting_down}
\end{eqnarray}
where in the above Eq.~(\ref{eq:creutz_point_splitting_down}),
the matrix $\Gamma$ is $\Gamma=i\gamma_4\gamma_5$, $\{\gamma_5,
\Gamma\}=0$ and $\Gamma^2=1$. The presence of the $\Gamma$ matrix
distinguishes the opposite chiralities of the two tastes. In the
configuration space, the Creutz point-splitting relations for KW
fermions are realized as,
\begin{eqnarray}
 u_x &=& \frac{1}{2}e^{i\frac{\pi}{2}x_4}\bigg(\psi_x+\frac{i}{2}(
\psi_{x-e_4}-\psi_{x+e_4})\bigg),
\label{eq:config_creutz_point_splitting_up}\\
 d_x &=& \frac{1}{2}\Gamma e^{-i\frac{\pi}{2}x_4}\bigg(\psi_x-
\frac{i}{2}(\psi_{x-e_4}-\psi_{x+e_4})\bigg).
\label{eq:config_creutz_point_splitting_down}
\end{eqnarray}
To write the free KW action in terms of the point-split tastes $u_x$
and $d_x$, we reorganize the Eqs.~(\ref{eq:config_creutz_point_splitting_up})
and (\ref{eq:config_creutz_point_splitting_down}) as,
\begin{eqnarray}
 \psi_x &=& e^{-i\frac{\pi}{2}x_4}u_x+e^{i\frac{\pi}{2}x_4}\Gamma d_x
\label{eq:inverse_point-splitting_psi}\\
 \overline{\psi}_x &=& e^{i\frac{\pi}{2}x_4}\overline{u}_x-e^{-i
\frac{\pi}{2}x_4}\overline{d}_x\Gamma
\label{eq:inverse_point-splitting_psibar}
\end{eqnarray}
and substitute the reorganized relations in the free KW action of
Eq.~(\ref{eq:free_lattice_action}). We further introduce a taste
isospin notation $\Psi$ in two-dimensional taste space as,
\begin{equation}
 \Psi_x=\begin{pmatrix} u_x\\ d_x\end{pmatrix} \;\;\text{and}
\;\; \overline{\Psi}_x=\begin{pmatrix} \overline{u}_x & \overline{d}_x
 \end{pmatrix}.\label{eq:isospinor}
\end{equation}
In terms of the taste isospinor $\Psi$, we get the KW point-split
action in configuration space,
\begin{equation}
\begin{split}
 \mathcal{S} = & \frac{1}{2} \sum\limits^3_{k=1} \sum\limits_x \biggl[
\begin{aligned}[t]
 & \overline{\Psi}_x \bigg\{(\gamma_k\otimes\sigma_3)+(- 1)^{x_4}(
\Gamma\gamma_k\otimes i\sigma_2) \bigg\} \Psi_{x+k} \\
 & -\overline{\Psi}_{x+k} \bigg\{(\gamma_k\otimes\sigma_3)+
(- 1)^{x_4}(\Gamma\gamma_k\otimes i\sigma_2) \bigg\} \Psi_{x} \biggr] 
\end{aligned} \\
 &+\frac{1}{2} \sum\limits_x \biggl[
\begin{aligned}[t]
 & \overline{\Psi}_x \bigg\{-(\gamma_4\otimes\sigma_3)- (- 1)^{x_4}(
\Gamma\gamma_4\otimes i\sigma_2) \bigg\} \Psi_{x+e_4} \\
 & +\overline{\Psi}_{x+e_4} \bigg\{(\gamma_4\otimes\sigma_3)-
(- 1)^{x_4}(\Gamma\gamma_4\otimes i\sigma_2) \bigg\}\Psi_{x} \biggr]
\end{aligned} \\
 &-\frac{i}{2} \sum\limits^3_{k=1} \sum\limits_x \biggl[
\begin{aligned}[t]
 & \overline{\Psi}_{x} \bigg\{(\gamma_4\otimes1)-(- 1)^{x_4}(\Gamma
\gamma_4\otimes \sigma_1) \bigg\} \Psi_{x+k} \\
 & +\overline{\Psi}_{x+k}\bigg\{(\gamma_4\otimes1)-(- 1)^{x_4}(
\Gamma\gamma_4\otimes \sigma_1) \bigg\} \Psi_{x} \\
 & -2\overline{\Psi}_{x}\bigg\{ (\gamma_4\otimes1)-(- 1)^{x_4}(\Gamma
\gamma_4\otimes \sigma_1) \bigg\} \Psi_{x} \biggr]
\end{aligned} \\
 & +m_0\sum\limits_x \biggl[ \overline{\Psi}_{x} \bigg\{(1\otimes
\sigma_3)+ (- 1)^{x_4}(\Gamma\otimes i\sigma_2) \bigg\} \Psi_{x}
\biggr] \label{eq:creutz_point-split_free_action}
\end{split}
\end{equation}
where the tensor product in the action indicates $\text{Dirac matrix}
\otimes\text{Taste matrix}$. In the order of their appearance in the
action, the terms can be understood as,
\begin{equation}
\begin{split}
\mathcal{S} = \text{[Kinetic Spatial]} + \text{[Kinetic Temporal]} \\
    +\text{[Karsten-Wilczek]}+\text{[Mass]} \label{eq:labeling_eqn}
\end{split}
\end{equation}
As an example, let us consider an explicit construction of one of
the terms, say the mass term in the last line of the above
point-split action Eq.~(\ref{eq:creutz_point-split_free_action}),
\begin{eqnarray}
 m_0 \sum_x \overline{\psi}_x\psi_x &=& m_0 \sum\limits_x
\Big[ e^{i\frac{\pi}{2}x_4} \overline{u}_x-e^{-i\frac{\pi}{2}x_4}
\overline{d}_x\Gamma \Big] \times \Big[ e^{-i\frac{\pi}{2}x_4}
u_x+e^{i\frac{\pi}{2}x_4}\Gamma d_x \Big]\nonumber\\
 &=& m_0 \sum_x \Big[ \overline{u}_xu_x+(- 1)^{x_4} \overline{u}_x
\Gamma d_x  -\;\overline{d}_xd_x-(- 1)^{x_4} \overline{d}_x \Gamma u_x
\Big] \nonumber\\
 &=& m_0 \sum_x \overline{\Psi}_{x} \Big\{ (1\otimes\sigma_3)+
(- 1)^{x_4}(\Gamma\otimes i\sigma_2) \Big\} \Psi_{x}
\label{eq:example_splitaction}
\end{eqnarray}
The remaining terms in the point-split action can be constructed
in a similar way.

The symmetries of the point-split action Eq.
(\ref{eq:creutz_point-split_free_action}) in terms of taste isospin
field $\Psi_x$ are given in Table \ref{tab:ptsplit_sym},
\begin{table*}
\caption{\label{tab:ptsplit_sym} Symmetries of point-split free action
Eq. (\ref{eq:creutz_point-split_free_action})}
\begin{ruledtabular}
\begin{tabular}{lll} 
1. Spatial Rotations \;\;\; & 2. Parity & 3. Charge conjugation $\times$
Time reversal \\
4. Site reflection \;\;\; & 5. $U(1)_\text{V}$ \;\;& \\
\end{tabular}
\end{ruledtabular}
\end{table*}
\noindent
Charge conjugation and Time reversal individually are not symmetries
of the original action but their product is. Charge conjugation and
Time reversal are broken by the Kinetic temporal and Karsten-Wilczek
term.

From the symmetry arguments, the taste isospinor $\Psi_x$ can transform
under four groups -- $U(1)_\text{V}$, $U(1)_\text{A}$, $SU(2)_\text{V}$
and $SU(2)_\text{A}$. Out of these only the $U(1)_\text{V}$ preserves
the taste isospin symmetry. The $SU(2)_\text{V}$ has three generators,
namely, $(1\otimes\sigma_1)$, $(1\otimes\sigma_2)$ and $(1\otimes
\sigma_3)$. Out of these, the Karsten-Wilczek term breaks $(1\otimes
\sigma_2)$ and $(1\otimes\sigma_3)$ but preserves $(1\otimes\sigma_1)$.
The remaining terms of the free action break all generators of
$SU(2)_\text{V}$.

The group $U(1)_\text{A}$ is generated by $(\gamma_5\otimes1)$. This
group is broken by all the terms of the action. The $SU(2)_\text{A}$
has three generators -- $(\gamma_5\otimes\sigma_1)$, $(\gamma_5\otimes
\sigma_2)$ and $(\gamma_5\otimes\sigma_3)$. The Kinetic spatial and
Kinetic temporal terms break $(\gamma_5\otimes\sigma_1)$ and $(\gamma_5
\otimes\sigma_2)$ but preserve the generator $(\gamma_5\otimes\sigma_3)$.
Whereas, the Mass term breaks $(\gamma_5\otimes \sigma_1)$ and $(
\gamma_5\otimes \sigma_3)$ but preserves  $(\gamma_5\otimes \sigma_2)$.
On the other hand, the KW term breaks $(\gamma_5\otimes \sigma_1)$
but preserves $(\gamma_5\otimes \sigma_2)$ and $(\gamma_5\otimes
\sigma_3)$. Note that in the chiral limit, $m_0\rightarrow0$, the
entire point-split action is invariant under the generator $(\gamma_5
\otimes \sigma_3)$ of $SU(2)_{\text{A}}$. The last line is crucial
for the construction of the Symanzik effective theory in the next
section.

A list of discrete transformations and the way they act on $\psi_x$
and $\Psi_x$ are given in the Table~\ref{tab:symm_transfrm}. Once
the symmetries of the action are made explicit, the next step is to
construct the free Symanzik effective action. We can now move to
the next step and using these symmetries, construct the free
Symanzik effective action.

\begin{table*}[hb]
\caption{\label{tab:symm_transfrm} Symmetry transformations and
their actions on fermionic $\psi_x$ and isospinor $\Psi_x$. The
first three transformations are linear while Site-reflection is
anti-linear. Charge Conjugation matrix $C=i\gamma_2\gamma_4$,
Time reversal matrix $T=\gamma_4\gamma_5$  and $\Gamma=i\gamma_4
\gamma_5$.}
\begin{ruledtabular}
\begin{tabular}{ccc}
 Transformation & Action on $\psi_x$ & Action on $\Psi_x$\\ \hline
 Parity & $\gamma_4\psi_{\mathbf{-x},x_4}$ & $(\gamma_4\otimes\sigma_3)
\Psi_{\mathbf{-x},x_4}$ \\
 Charge Conjugation & $C^{- 1}\overline{\psi}_x^T$ & $(C^{- 1}\Gamma\otimes
i\sigma_2)\overline{\Psi}^T_x$\\
 Time Reversal & $T\psi_{\mathbf{x},-x_4}$ & $(T\Gamma\otimes\sigma_1)
\Psi_{\mathbf{x},-x_4}$\\
 Site Reflection & $T\overline{\psi}^T_{\mathbf{1-x},x_4}$ & $(T\otimes
\sigma_3)\overline{\Psi}^T_{\mathbf{1-x},x_4}$\\
\end{tabular}
\end{ruledtabular}
\end{table*}

\section{Free Symanzik Effective Action for KW Fermions
\label{sec:free_symanzik_action}}

We construct the Symanzik effective action to NLO by considering
operators up to dimension-5. The Symanzik action has the following
form in terms of lattice spacing $a$,
\begin{equation}
\mathcal{S}_\text{Symanzik} =a^{- 1}S_{- 1}+S_0+aS_1+a^2S_2+...
\label{eq:symanzik_def_1}
\end{equation}
where, $S_k$ consists of operators of dimension-$(k+4)$. For
instance, for dimension-3 operators, $k=- 1$, we have the term
$a^{- 1}S_{- 1}$. The operators that go in $S_k$ are invariant
under the symmetries of the lattice action, which in this case
is the point-split KW action in Eq.
(\ref{eq:creutz_point-split_free_action}). Corresponding to
$l$-different operators of dimension-$(k+4)$, we associate with
$\mathcal{O}^{(k+4)}_{l}(x)$ a coefficient $c_l^{(k+4)}$, often
referred to as the low-energy constant (LEC). 
\begin{equation}
S_k=\sum\limits_l\int d^4x\, c_l^{(k+4)}\mathcal{O}_l^{(k+4)}
\label{eq:symanzik_operator}
\end{equation}
We tabulate the $(k+4)$-dimension operators up to $k=1$ in the
Tables~\ref{tab:symanzik_dimension-3},\,\ref{tab:symanzik_dimension-4a},\,
\ref{tab:symanzik_dimension-4b},\, \ref{tab:symanzik_dimension-5a},\,
\ref{tab:symanzik_dimension-5b} and \ref{tab:symanzik_dimension-5c}. In these
tables, the leftmost column lists the rotationally invariant
operators of appropriate dimensions. The following columns show
the invariance property or lack of it for various symmetry
transformations, except for the second column which discusses the
invariance property under the generator $(\gamma_5\otimes\sigma_3)$
of the group $SU(2)_A$ in the taste space. The $\mathcal{I}$ entries in a
row symbolize invariance of the operator under such transformations
and the $\times$ entries symbolize the breaking of the invariance. The
rows with single entries of $\mathcal{I}$ or $\times$ for every symmetry
transformation correspond to operators with unit matrix in the
taste space while the rows with three entries correspond to operators
having one of the three $\sigma_1$, $\sigma_2$, $\sigma_3$  in
the taste space.

None of the tables here contain dimension-2 operators since no
local operators of dimension-2 that are invariant under rotations
and gauge transformations can be constructed. Hence the dimension-3
operators are the LO operators in the Symanzik action Eq.
(\ref{eq:symanzik_operator}).

\begin{table*}[h]
\caption{\label{tab:symanzik_dimension-3} The rotationally invariant
dimension-3 operators and their transformations under various
symmetries and a generator of $SU(2)_A$. Invariance of an operator
is denoted by $\mathcal{I}$ while those that are not are denoted by $\times$.}
\begin{ruledtabular}
\begin{tabular}{ccccccc}
Operators & $(\gamma_5\otimes\sigma_3)$ & Parity &
Time & Charge & Charge$\times$Time & Site \\
 & of $SU(2)_\text{A}$ & & Reversal & Conjugation & & Reflection \\ \hline    
$\overline{\Psi}(1\otimes1)\Psi$ & $\times$ & $\mathcal{I}$ & $\times$ & $\times$ &$\mathcal{I}$& $\mathcal{I}$\\
$\overline{\Psi}(1\otimes\sigma_i)\Psi$ & $\mathcal{I}$, $\mathcal{I}$, $\times$  &$\times$,$\times$,$\mathcal{I}$ &
$\times$, $\mathcal{I}$, $\mathcal{I}$ & $\mathcal{I}$, $\mathcal{I}$, $\mathcal{I}$ &$\times$, $\mathcal{I}$, $\mathcal{I}$ & $\times$, $\times$, $\mathcal{I}$\\ \hline
$\overline{\Psi}(\gamma_4\otimes1)\Psi$ & $\mathcal{I}$ &$\mathcal{I}$ & $\times$ & $\times$ &$\mathcal{I}$& $\times$\\
$\overline{\Psi}(\gamma_4\otimes\sigma_i)\Psi$ & $\times$, $\times$, $\mathcal{I}$ &$\times$,$\times$,$\mathcal{I}$
& $\times$, $\mathcal{I}$, $\mathcal{I}$ & $\mathcal{I}$, $\mathcal{I}$, $\mathcal{I}$ &$\times$, $\mathcal{I}$, $\mathcal{I}$ & $\mathcal{I}$, $\mathcal{I}$, $\times$\\ \hline
$\overline{\Psi}(\gamma_5\otimes1)\Psi$ & $\times$ &$\times$ & $\times$ & $\mathcal{I}$& $\times$ & $\times$\\
$\overline{\Psi}(\gamma_5\otimes\sigma_i)\Psi$ & $\mathcal{I}$, $\mathcal{I}$, $\times$ &$\mathcal{I}$,$\mathcal{I}$,$\times$
& $\times$, $\mathcal{I}$, $\mathcal{I}$ & $\times$, $\times$, $\times$ &$\mathcal{I}$, $\times$, $\times$ & $\mathcal{I}$, $\mathcal{I}$, $\times$\\ \hline
$\overline{\Psi}(\gamma_5\gamma_4\otimes1)\Psi$ & $\mathcal{I}$ &$\times$ & $\times$ & $\times$
&$\mathcal{I}$ & $\times$\\
$\overline{\Psi}(\gamma_5\gamma_4\otimes\sigma_i)\Psi$& $\times$, $\times$, $\mathcal{I}$
&$\mathcal{I}$,$\mathcal{I}$,$\times$ & $\times$, $\mathcal{I}$, $\mathcal{I}$ & $\mathcal{I}$, $\mathcal{I}$, $\mathcal{I}$ &$\times$, $\mathcal{I}$, $\mathcal{I}$ & $\mathcal{I}$, $\mathcal{I}$, $\times$\\
\end{tabular}
\end{ruledtabular}
\end{table*}
In Table \ref{tab:symanzik_dimension-3}, the transformation properties of
dimension-3 operators are listed. From the table, as an example, we
consider the fourth row which shows the invariance properties of
the operator $\overline{\Psi}(\gamma_4\otimes\sigma_i)\Psi$. We
check the transformation of this term under parity, for instance.
Because this terms corresponds to the taste matrix $\sigma_i$, the
parity invariance holds only for $\sigma_3$ hence $\mathcal{I}$. Whereas for
the two other sigma matrices, parity invariance is broken, hence
$\times$. Because parity invariance is the symmetry of the lattice action,
operators entering the Symanzik action are required to satisfy parity
invariance as well. Hence, the operators $\overline{\Psi}(\gamma_4
\otimes\sigma_i)\Psi$ with $i=1,2$ cannot be included in the Symanzik
action. In Appendix A, we provide details for calculating the
symmetry properties of the operator taking a dimension-4 operator,
appearing in Table \ref{tab:symanzik_dimension-4a} as an example.
From the dimension-3 operators given in Table~\ref{tab:symanzik_dimension-3},
based on the discussion above, the operators that satisfy all the
symmetries of the lattice action qualify for the Symanzik action.
They are, 
\begin{align}
i\overline{\Psi}(\gamma_4\otimes1)\Psi, \qquad i\overline{\Psi}(\gamma_4
\otimes\sigma_3)\Psi, \qquad\overline{\Psi}(\gamma_5\otimes\sigma_1)
\Psi.\label{eq:symanzik_dimension-3_free}
\end{align}
The Tables~\ref{tab:symanzik_dimension-4a} and \ref{tab:symanzik_dimension-4b}
list all rotationally invariant dimension-4 operators with single
derivatives $(\partial_k, \partial_4)$ and one power of the quark
mass $m$, respectively.

\begin{table*}[h]
\caption{\label{tab:symanzik_dimension-4a}The rotationally invariant
dimension-4 operators containing a partial derivative.}
\begin{ruledtabular}
\begin{tabular}{ccccccc}
Operators & $(\gamma_5\otimes\sigma_3)$ & Parity &
Time & Charge & Charge$\times$Time & Site \\
 & of $SU(2)_\text{A}$ & & Reversal & Conjugation & & Reflection \\ \hline    
$\overline{\Psi}(1\otimes1)\partial_4\Psi$ & $\times$ & $\mathcal{I}$& $\mathcal{I}$ & $\mathcal{I}$ &$\mathcal{I}$ & $\times $ \\
$\overline{\Psi}(1\otimes\sigma_i)\partial_4\Psi$& $\mathcal{I}$, $\mathcal{I}$, $\times$
&$\times$, $\times$, $\mathcal{I}$ & $\mathcal{I}$, $\times$, $\times$ & $\times$, $\times$, $\times$ &$\times$, $\mathcal{I}$, $\mathcal{I}$ & $\mathcal{I}$, $\mathcal{I}$, $\times$\\ \hline
$\overline{\Psi}(\gamma_4\otimes1)\partial_4\Psi$& $\mathcal{I}$ &$\mathcal{I}$ & $\mathcal{I}$ & $\mathcal{I}$ &
$\mathcal{I}$ & $\mathcal{I}$ \\
$\overline{\Psi}(\gamma_4\otimes\sigma_i)\partial_4\Psi$& $\times$, $\times$, $\mathcal{I}$
&$\times$, $\times$, $\mathcal{I}$ & $\mathcal{I}$, $\times$, $\times$ & $\times$, $\times$, $\times$ &$\times$, $\mathcal{I}$, $\mathcal{I}$ & $\times$, $\times$, $\mathcal{I}$ \\ \hline
$\overline{\Psi}(\gamma_5\otimes1)\partial_4\Psi$& $\times$ &$\times$ & $\mathcal{I}$ 
& $\times$& $\times$ & $\mathcal{I}$ \\
$\overline{\Psi}(\gamma_5\otimes\sigma_i)\partial_4\Psi$&$\mathcal{I}$, $\mathcal{I}$, $\times$
&$\mathcal{I}$, $\mathcal{I}$, $\times$ & $\mathcal{I}$, $\times$, $\times$ & $\mathcal{I}$, $\mathcal{I}$, $\mathcal{I}$ & $\mathcal{I}$, $\times$, $\times$& $\times$, $\times$, $\mathcal{I}$ \\ \hline
$\overline{\Psi}(\gamma_5\gamma_4\otimes1)\partial_4\Psi$&$\mathcal{I}$  &$\times$
& $\mathcal{I}$ & $\mathcal{I}$& $\mathcal{I}$ & $\mathcal{I}$ \\
$\overline{\Psi}(\gamma_5\gamma_4\otimes\sigma_i)\partial_4\Psi$&$\times$, $\times$,
$\mathcal{I}$  &$\mathcal{I}$, $\mathcal{I}$, $\times$ & $\mathcal{I}$, $\times$, $\times$ & $\times$, $\times$, $\times$ &$\times$, $\mathcal{I}$, $\mathcal{I}$ & $\times$, $\times$, $\mathcal{I}$ \\ \hline
$\overline{\Psi}(\gamma_k\otimes1)\partial_k\Psi$& $\mathcal{I}$ & $\mathcal{I}$& $\times$
& $\times$ & $\mathcal{I}$& $\mathcal{I}$ \\
$\overline{\Psi}(\gamma_k\otimes\sigma_i)\partial_k\Psi$& $\times$,
$\times$, $\mathcal{I}$ &$\times$, $\times$, $\mathcal{I}$ & $\times$, $\mathcal{I}$, $\mathcal{I}$ & $\mathcal{I}$, $\mathcal{I}$, $\mathcal{I}$& $\times$, $\mathcal{I}$, $\mathcal{I}$ & $\times$, $\times$, $\mathcal{I}$ \\
\hline
$\overline{\Psi}(\gamma_5\gamma_k\otimes1)\partial_k\Psi$&$\mathcal{I}$
&$\times$& $\times$ & $\times$ & $\mathcal{I}$& $\mathcal{I}$ \\
$\overline{\Psi}(\gamma_5\gamma_k\otimes\sigma_i)\partial_k\Psi$&
$\times$, $\times$, $\mathcal{I}$ &$\mathcal{I}$, $\mathcal{I}$, $\times$ & $\times$, $\mathcal{I}$, $\mathcal{I}$ & $\mathcal{I}$, $\mathcal{I}$, $\mathcal{I}$ & $\times$, $\mathcal{I}$, $\mathcal{I}$& $\times$,
$\times$, $\mathcal{I}$ \\ \hline
$\overline{\Psi}(\gamma_4\gamma_k\otimes 1)\partial_k\Psi$ &
$\times$ & $\mathcal{I}$ & $\times$ & $\mathcal{I}$ &$\times$ & $\mathcal{I}$ \\
$\overline{\Psi}(\gamma_4\gamma_k\otimes \sigma_i)\partial_k\Psi$
& $\mathcal{I}$, $\mathcal{I}$, $\times$ & $\times$, $\times$, $\mathcal{I}$ & $\times$, $\mathcal{I}$, $\mathcal{I}$ & $\times$, $\times$, $\times$ &$\mathcal{I}$, $\times$, $\times$ & $\times$,
$\times$, $\mathcal{I}$ \\
\end{tabular}
\end{ruledtabular}
\end{table*}

\begin{table*}[h]
\caption{\label{tab:symanzik_dimension-4b} The rotationally invariant
dimension-4 operators proportional to the fermion mass $m$.}
\begin{ruledtabular}
\begin{tabular}{ccccccc}
Operators & $(\gamma_5\otimes\sigma_3)$ & Parity &
Time & Charge & Charge$\times$Time & Site \\
 & of $SU(2)_\text{A}$ & & Reversal & Conjugation & & Reflection \\ \hline    
$m\overline{\Psi}(1\otimes1)\Psi$ & $\times$ & $\mathcal{I}$ & $\times$ & $\times$ & $\mathcal{I}$& $\mathcal{I}$\\
$m\overline{\Psi}(1\otimes\sigma_i)\Psi$ & $\mathcal{I}$, $\mathcal{I}$, $\times$  &$\times$,$\times$,$\mathcal{I}$ &
$\times$, $\mathcal{I}$, $\mathcal{I}$ & $\mathcal{I}$, $\mathcal{I}$, $\mathcal{I}$ & $\times$, $\mathcal{I}$, $\mathcal{I}$& $\times$, $\times$, $\mathcal{I}$\\ \hline
$m\overline{\Psi}(\gamma_4\otimes1)\Psi$ & $\mathcal{I}$ &$\mathcal{I}$ & $\times$ & $\times$ &$\mathcal{I}$ & $\times$\\
$m\overline{\Psi}(\gamma_4\otimes\sigma_i)\Psi $ & $\times$, $\times$, $\mathcal{I}$ &$\times$,$\times$,
$\mathcal{I}$ & $\times$, $\mathcal{I}$, $\mathcal{I}$ & $\mathcal{I}$, $\mathcal{I}$, $\mathcal{I}$ &$\times$, $\mathcal{I}$, $\mathcal{I}$ & $\mathcal{I}$, $\mathcal{I}$, $\times$\\ \hline
$m\overline{\Psi}(\gamma_5\otimes1)\Psi$ & $\times$ &$\times$ & $\times$ & $\mathcal{I}$ & $\times$& $\times$\\
$m\overline{\Psi}(\gamma_5\otimes\sigma_i)\Psi$ & $\mathcal{I}$, $\mathcal{I}$, $\times$ &$\mathcal{I}$,$\mathcal{I}$,$\times$
& $\times$, $\mathcal{I}$, $\mathcal{I}$ & $\times$, $\times$, $\times$& $\mathcal{I}$, $\times$, $\times$ & $\mathcal{I}$, $\mathcal{I}$, $\times$\\ \hline
$m\overline{\Psi}(\gamma_5\gamma_4\otimes1)\Psi$ & $\mathcal{I}$ &$\times$ & $\times$ & $\times$
&$\mathcal{I}$ & $\times$\\
$m\overline{\Psi}(\gamma_5\gamma_4\otimes\sigma_i)\Psi$& $\times$, $\times$, $\mathcal{I}$
&$\mathcal{I}$,$\mathcal{I}$,$\times$ & $\times$, $\mathcal{I}$, $\mathcal{I}$ & $\mathcal{I}$, $\mathcal{I}$, $\mathcal{I}$ &$\times$, $\mathcal{I}$, $\mathcal{I}$ & $\mathcal{I}$, $\mathcal{I}$, $\times$\\ \hline
\end{tabular}
\end{ruledtabular}
\end{table*}
From a similar reasoning to one provided in selection of dimension-3
operator, the terms that enter the Symanzik action in NLO are,
where sum over $k$ is understood,
\begin{align}
 & \overline{\Psi}(\gamma_4\otimes1)\partial_4\Psi, && \overline{\Psi}
(\gamma_4\otimes\sigma_3)\partial_4\Psi, && i\overline{\Psi}(\gamma_5
\otimes\sigma_1)\partial_4\Psi, && \overline{\Psi}(\gamma_k\otimes1)
\partial_k\Psi, \nonumber \\
 & \overline{\Psi}(\gamma_k\otimes\sigma_3)\partial_k\Psi, &&
m\overline{\Psi} (1\otimes1)\Psi, && m\overline{\Psi}(1\otimes\sigma_3)
\Psi, && im\overline{\Psi}(\gamma_4\otimes1)\Psi, \nonumber \\
 & im\overline{\Psi}(\gamma_4\otimes\sigma_3)\Psi, && m\overline{
\Psi}(\gamma_5\otimes\sigma_1)\Psi, && m\overline{\Psi}
(\gamma_5\gamma_4\otimes\sigma_2)\Psi && \label{dimension3op}
\end{align}

\begin{table*}[h]
\caption{\label{tab:symanzik_dimension-5a} The rotationally invariant
dimension-5 operators containing two partial derivatives.}
\begin{ruledtabular}
\begin{tabular}{ccccccc}
Operators & $(\gamma_5\otimes\sigma_3)$ & Parity &
Time & Charge & Charge$\times$Time & Site \\
 & of $SU(2)_\text{A}$ & & Reversal & Conjugation & & Reflection \\ \hline    
 $\overline{\Psi}(1\otimes1)\partial_4\partial_4\Psi$ & $\times$ & $\mathcal{I}$ & $\times$
& $\times$ & $\mathcal{I}$& $\mathcal{I}$ \\
 $\overline{\Psi}(1\otimes\sigma_i)\partial_4\partial_4\Psi$ & $\mathcal{I}$, $\mathcal{I}$,
$\times$  & $\times$,$\times$,$\mathcal{I}$ & $\times$,$\mathcal{I}$,$\mathcal{I}$ & $\mathcal{I}$,$\mathcal{I}$,$\mathcal{I}$ & $\times$,$\mathcal{I}$,$\mathcal{I}$ & $\times$,$\times$,$\mathcal{I}$ \\ \hline
 $\overline{\Psi}(\gamma_4\otimes1)\partial_4\partial_4\Psi$ & $\mathcal{I}$ & $\mathcal{I}$
& $\times$ & $\times$& $\mathcal{I}$ & $\times$ \\
 $\overline{\Psi}(\gamma_4\otimes\sigma_i)\partial_4\partial_4\Psi$
 & $\times$, $\times$, $\mathcal{I}$ & $\times$, $\times$, $\mathcal{I}$ & $\times$, $\mathcal{I}$, $\mathcal{I}$ & $\mathcal{I}$, $\mathcal{I}$, $\mathcal{I}$ &$\times$, $\mathcal{I}$, $\mathcal{I}$ &
$\mathcal{I}$, $\mathcal{I}$, $\times$ \\ \hline
 $\overline{\Psi}(\gamma_5\otimes1)\partial_4\partial_4\Psi$ & $\times$ &
$\times$ & $\times$ & $\mathcal{I}$ & $\times$& $\times$ \\
 $\overline{\Psi}(\gamma_5\otimes\sigma_i)\partial_4\partial_4\Psi$ &
$\mathcal{I}$, $\mathcal{I}$, $\times$ & $\mathcal{I}$, $\mathcal{I}$, $\times$ & $\times$, $\mathcal{I}$, $\mathcal{I}$ & $\times$, $\times$, $\times$& $\mathcal{I}$, $\times$, $\times$ & $\mathcal{I}$,
$\mathcal{I}$, $\times$ \\ \hline
 $\overline{\Psi}(\gamma_5\gamma_4\otimes1)\partial_4\partial_4\Psi$
& $\mathcal{I}$ & $\times$ & $\times$ & $\times$ & $\mathcal{I}$& $\times$ \\
 $\overline{\Psi}(\gamma_5\gamma_4\otimes\sigma_i)\partial_4\partial_4
\Psi$ & $\times$, $\times$, $\mathcal{I}$ & $\mathcal{I}$, $\mathcal{I}$, $\times$ & $\times$, $\mathcal{I}$, $\mathcal{I}$ & $\mathcal{I}$, $\mathcal{I}$, $\mathcal{I}$& $\times$, $\mathcal{I}$, $\mathcal{I}$
 & $\mathcal{I}$, $\mathcal{I}$, $\times$ \\ \hline
 $\overline{\Psi}(1\otimes1)\partial_k\partial_k\Psi$ & $\times$ & $\mathcal{I}$ & $\times$
 & $\times$ & $\mathcal{I}$& $\mathcal{I}$ \\
 $\overline{\Psi}(1\otimes\sigma_i)\partial_k\partial_k\Psi$ & $\mathcal{I}$, $\mathcal{I}$,
$\times$ & $\times$,$\times$,$\mathcal{I}$ & $\times$,$\mathcal{I}$,$\mathcal{I}$ & $\mathcal{I}$,$\mathcal{I}$,$\mathcal{I}$ & $\times$,$\mathcal{I}$,$\mathcal{I}$ & $\times$,$\times$,$\mathcal{I}$ \\ \hline
 $\overline{\Psi}(\gamma_4\otimes1)\partial_k\partial_k\Psi$ & $\mathcal{I}$ &
$\mathcal{I}$ & $\times$ & $\times$& $\mathcal{I}$ & $\times$ \\
 $\sum_k\overline{\Psi}(\gamma_4\otimes\sigma_i)\partial_k\partial_k
\Psi$ & $\times$, $\times$, $\mathcal{I}$ & $\times$, $\times$, $\mathcal{I}$ & $\times$, $\mathcal{I}$, $\mathcal{I}$ & $\mathcal{I}$, $\mathcal{I}$, $\mathcal{I}$& $\times$, $\mathcal{I}$, $\mathcal{I}$
& $\mathcal{I}$, $\mathcal{I}$, $\times$ \\ \hline
 $\overline{\Psi}(\gamma_5\otimes1)\partial_k\partial_k\Psi$ & $\times$ &
$\times$ & $\times$ & $\mathcal{I}$& $\times$ & $\times$ \\
 $\overline{\Psi}(\gamma_5\otimes\sigma_i)\partial_k\partial_k\Psi$ &
$\mathcal{I}$, $\mathcal{I}$, $\times$ & $\mathcal{I}$, $\mathcal{I}$, $\times$ & $\times$, $\mathcal{I}$, $\mathcal{I}$ & $\times$, $\times$, $\times$& $\mathcal{I}$, $\times$, $\times$ & $\mathcal{I}$,
$\mathcal{I}$, $\times$ \\ \hline
 $\overline{\Psi}(\gamma_5\gamma_4\otimes1)\partial_k\partial_k\Psi$
 & $\mathcal{I}$ & $\times$ & $\times$ & $\times$& $\mathcal{I}$ & $\times$ \\
 $\overline{\Psi}(\gamma_5\gamma_4\otimes\sigma_i)\partial_k
\partial_k\Psi$& $\times$, $\times$, $\mathcal{I}$ & $\mathcal{I}$, $\mathcal{I}$, $\times$ & $\times$, $\mathcal{I}$, $\mathcal{I}$ & $\mathcal{I}$, $\mathcal{I}$, $\mathcal{I}$
& $\times$, $\mathcal{I}$, $\mathcal{I}$ & $\mathcal{I}$, $\mathcal{I}$, $\times$\\ \hline
 $\overline{\Psi}(\gamma_k\otimes1)\partial_4\partial_k\Psi$ & $\mathcal{I}$ &
$\mathcal{I}$ & $\mathcal{I}$ & $\mathcal{I}$& $\mathcal{I}$ & $\times$ \\
 $\overline{\Psi}(\gamma_k\otimes\sigma_i)\partial_4\partial_k\Psi$
& $\times$, $\times$, $\mathcal{I}$ & $\times$, $\times$, $\mathcal{I}$ & $\mathcal{I}$, $\times$, $\times$ & $\times$, $\times$, $\times$& $\times$, $\mathcal{I}$, $\mathcal{I}$
& $\mathcal{I}$, $\mathcal{I}$, $\times$ \\ \hline
 $\overline{\Psi}(\gamma_4\gamma_k\otimes1)\partial_4\partial_k\Psi$
& $\times$ & $\mathcal{I}$ & $\mathcal{I}$ & $\times$ & $\times$& $\times$ \\
 $\overline{\Psi}(\gamma_4\gamma_k\otimes\sigma_i)\partial_4\partial_k
\Psi$ & $\mathcal{I}$, $\mathcal{I}$, $\times$ & $\times$, $\times$, $\mathcal{I}$ & $\mathcal{I}$, $\times$, $\times$ & $\mathcal{I}$, $\mathcal{I}$, $\mathcal{I}$& $\mathcal{I}$, $\times$, $\times$
& $\mathcal{I}$, $\mathcal{I}$, $\times$ \\ \hline
 $\overline{\Psi}(\gamma_5\gamma_k\otimes1)\partial_4\partial_k\Psi$
& $\mathcal{I}$ & $\times$ & $\mathcal{I}$ & $\mathcal{I}$ &$\mathcal{I}$ & $\times$ \\
 $\overline{\Psi}(\gamma_5\gamma_k\otimes\sigma_i)\partial_4\partial_k
\Psi$ & $\times$, $\times$, $\mathcal{I}$ & $\mathcal{I}$, $\mathcal{I}$, $\times$ & $\mathcal{I}$, $\times$, $\times$ & $\times$, $\times$, $\times$& $\times$, $\mathcal{I}$, $\mathcal{I}$
& $\mathcal{I}$, $\mathcal{I}$, $\times$ \\
\end{tabular}
\end{ruledtabular}
\end{table*}

\begin{table*}[h]
\caption{\label{tab:symanzik_dimension-5b} The rotationally invariant
dimension-5 operators containing a partial derivative and proportional
to the fermion mass $m$. Sum over $k$ is assumed.}
\begin{ruledtabular}
\begin{tabular}{ccccccc}
Operators & $(\gamma_5\otimes\sigma_3)$ & Parity &
Time & Charge & Charge$\times$Time & Site \\
 & of $SU(2)_\text{A}$ & & Reversal & Conjugation & & Reflection \\ \hline    
$m\:\overline{\Psi}(1\otimes1)\partial_4\Psi$& $\times$ & $\mathcal{I}$& $\mathcal{I}$ & $\mathcal{I}$ &$\mathcal{I}$ & $\times$ \\
$m\:\overline{\Psi}(1\otimes\sigma_i)\partial_4\Psi$
& $\mathcal{I}$, $\mathcal{I}$, $\times$ &$\times$, $\times$, $\mathcal{I}$ & $\mathcal{I}$, $\times$, $\times$ & $\times$, $\times$, $\times$ &$\times$, $\mathcal{I}$, $\mathcal{I}$ 
& $\mathcal{I}$, $\mathcal{I}$, $\times$\\ \hline
$m\:\overline{\Psi}(\gamma_4\otimes1)\partial_4\Psi$
& $\mathcal{I}$ &$\mathcal{I}$ & $\mathcal{I}$ & $\mathcal{I}$ & $\mathcal{I}$ & $\mathcal{I}$ \\
$m\: \overline{\Psi}(\gamma_4\otimes\sigma_i)\partial_4\Psi$
& $\times$, $\times$, $\mathcal{I}$ &$\times$, $\times$, $\mathcal{I}$ & $\mathcal{I}$, $\times$, $\times$ & $\times$, $\times$, $\times$ &$\times$, $\mathcal{I}$, $\mathcal{I}$ 
& $\times$, $\times$, $\mathcal{I}$ \\ \hline
$m\:\overline{\Psi}(\gamma_5\otimes1)\partial_4\Psi$
& $\times$ &$\times$ & $\mathcal{I}$ & $\times$& $\times$ & $\mathcal{I}$ \\
$m\:\overline{\Psi}(\gamma_5\otimes\sigma_i)\partial_4\Psi$
&$\mathcal{I}$, $\mathcal{I}$, $\times$  &$\mathcal{I}$, $\mathcal{I}$, $\times$ & $\mathcal{I}$, $\times$, $\times$ & $\mathcal{I}$, $\mathcal{I}$, $\mathcal{I}$ & $\mathcal{I}$, $\times$, $\times$
& $\times$, $\times$, $\mathcal{I}$ \\ \hline
$m\: \overline{\Psi}(\gamma_5\gamma_4\otimes1)\partial_4\Psi$
&$\mathcal{I}$  &$\times$ & $\mathcal{I}$ & $\mathcal{I}$& $\mathcal{I}$ & $\mathcal{I}$ \\
$m\:\overline{\Psi}(\gamma_5\gamma_4\otimes\sigma_i)\partial_4\Psi$
&$\times$, $\times$, $\mathcal{I}$  &$\mathcal{I}$, $\mathcal{I}$, $\times$ & $\mathcal{I}$, $\times$, $\times$ & $\times$, $\times$, $\times$ &$\times$, $\mathcal{I}$, $\mathcal{I}$ & $\times$,
$\times$, $\mathcal{I}$ \\ \hline
$m\:\overline{\Psi}(\gamma_k\otimes1)\partial_k\Psi$
& $\mathcal{I}$ & $\mathcal{I}$& $\times$ & $\times$ & $\mathcal{I}$& $\mathcal{I}$ \\
$m\:\overline{\Psi}(\gamma_k\otimes\sigma_i)\partial_k\Psi$
& $\times$, $\times$, $\mathcal{I}$ &$\times$, $\times$, $\mathcal{I}$ & $\times$, $\mathcal{I}$, $\mathcal{I}$ & $\mathcal{I}$, $\mathcal{I}$, $\mathcal{I}$& $\times$, $\mathcal{I}$, $\mathcal{I}$ & $\times$,
$\times$, $\mathcal{I}$ \\ \hline
$m\:\overline{\Psi}(\gamma_5\gamma_k\otimes1)\partial_k\Psi$
&$\mathcal{I}$  &$\times$& $\times$ & $\times$ & $\mathcal{I}$& $\mathcal{I}$ \\
$m\:\overline{\Psi}(\gamma_5\gamma_k\otimes\sigma_i)\partial_k\Psi$
& $\times$, $\times$, $\mathcal{I}$ &$\mathcal{I}$, $\mathcal{I}$, $\times$ & $\times$, $\mathcal{I}$, $\mathcal{I}$ & $\mathcal{I}$, $\mathcal{I}$, $\mathcal{I}$ & $\times$, $\mathcal{I}$, $\mathcal{I}$
& $\times$, $\times$, $\mathcal{I}$ \\ \hline
$m\:\overline{\Psi}(\gamma_4\gamma_k\otimes 1)\partial_k\Psi$ & $\times$ 
& $\mathcal{I}$ & $\times$ & $\mathcal{I}$ &$\times$ & $\mathcal{I}$ \\
$m\:\overline{\Psi}(\gamma_4\gamma_k\otimes \sigma_i)\partial_k\Psi$
& $\mathcal{I}$, $\mathcal{I}$, $\times$ & $\times$, $\times$, $\mathcal{I}$ & $\times$, $\mathcal{I}$, $\mathcal{I}$ & $\times$, $\times$, $\times$ &$\mathcal{I}$, $\times$, $\times$ 
& $\times$, $\times$, $\mathcal{I}$ \\
\end{tabular}
\end{ruledtabular}
\end{table*}

\begin{table*}[h]
\caption{\label{tab:symanzik_dimension-5c} The rotationally invariant
dimension-5 operators proportional to $m^2$.}
\begin{ruledtabular}
\begin{tabular}{ccccccc}
Operators & $(\gamma_5\otimes\sigma_3)$ & Parity &
Time & Charge & Charge$\times$Time & Site \\
 & of $SU(2)_\text{A}$ & & Reversal & Conjugation & & Reflection \\ \hline    
$m^2\:\overline{\Psi}(1\otimes1)\Psi$ & $\times$ & $\mathcal{I}$ & $\times$ & $\times$ & $\mathcal{I}$& $\mathcal{I}$\\
$m^2\:\overline{\Psi}(1\otimes\sigma_i)\Psi$ & $\mathcal{I}$, $\mathcal{I}$, $\times$  &$\times$,$\times$,$\mathcal{I}$ 
& $\times$, $\mathcal{I}$, $\mathcal{I}$ & $\mathcal{I}$, $\mathcal{I}$, $\mathcal{I}$ & $\times$, $\mathcal{I}$, $\mathcal{I}$& $\times$, $\times$, $\mathcal{I}$\\ \hline
$m^2\:\overline{\Psi}(\gamma_4\otimes1)\Psi$ & $\mathcal{I}$ &$\mathcal{I}$ & $\times$ & $\times$ 
&$\mathcal{I}$ & $\times$\\
$m^2\:\overline{\Psi}(\gamma_4\otimes\sigma_i)\Psi$ & $\times$, $\times$, $\mathcal{I}$ 
&$\times$,$\times$,$\mathcal{I}$ & $\times$, $\mathcal{I}$, $\mathcal{I}$ & $\mathcal{I}$, $\mathcal{I}$, $\mathcal{I}$ &$\times$, $\mathcal{I}$, $\mathcal{I}$ & $\mathcal{I}$, $\mathcal{I}$, $\times$\\ \hline
$m^2\:\overline{\Psi}(\gamma_5\otimes1)\Psi$ & $\times$ &$\times$ & $\times$ & $\mathcal{I}$ 
& $\times$& $\times$\\
$m^2\:\overline{\Psi}(\gamma_5\otimes\sigma_i)\Psi$ & $\mathcal{I}$, $\mathcal{I}$, $\times$ 
&$\mathcal{I}$,$\mathcal{I}$,$\times$ & $\times$, $\mathcal{I}$, $\mathcal{I}$ & $\times$, $\times$, $\times$& $\mathcal{I}$, $\times$, $\times$ & $\mathcal{I}$, $\mathcal{I}$, $\times$\\ \hline
$m^2\:\overline{\Psi}(\gamma_5\gamma_4\otimes1)\Psi$ & $\mathcal{I}$ &$\times$ 
& $\times$ & $\times$ &$\mathcal{I}$ & $\times$\\
$m^2\:\overline{\Psi}(\gamma_5\gamma_4\otimes\sigma_i)\Psi$
& $\times$, $\times$, $\mathcal{I}$ &$\mathcal{I}$,$\mathcal{I}$,$\times$ & $\times$, $\mathcal{I}$, $\mathcal{I}$ & $\mathcal{I}$, $\mathcal{I}$, $\mathcal{I}$ &$\times$, $\mathcal{I}$, $\mathcal{I}$ 
& $\mathcal{I}$, $\mathcal{I}$, $\times$\\
\end{tabular}
\end{ruledtabular}
\end{table*}

Similarly, the operators coming from Tables~\ref{tab:symanzik_dimension-5a},
\ref{tab:symanzik_dimension-5b} and \ref{tab:symanzik_dimension-5c} that enter
the NNLO of the Symanzik action are,
\begin{align}
 & i\overline{\Psi}(\gamma_4\otimes1)\partial_4\partial_4\Psi, &&
i\overline{\Psi}(\gamma_4\otimes\sigma_3)\partial_4\partial_4\Psi, &&
\overline{\Psi}(\gamma_5\otimes\sigma_1)\partial_4\partial_4\Psi, &&
 i\overline{\Psi}(\gamma_4\otimes1)\partial_k\partial_k\Psi, \nonumber \\
 & i\overline{\Psi}(\gamma_4\otimes\sigma_3)\partial_k\partial_k\Psi,
 && \overline{\Psi}(\gamma_5\otimes\sigma_1)\partial_k\partial_k\Psi,
 && i\overline{\Psi}(\gamma_k\otimes1)\partial_4\partial_k\Psi, &&
 i\overline{\Psi}(\gamma_k\otimes\sigma_3)\partial_4\partial_k\Psi,
 \nonumber \\
 & m\overline{\Psi}(\gamma_4\otimes1)\partial_4\Psi, &&
 m\overline{\Psi}(\gamma_4\otimes\sigma_3)\partial_4\Psi, &&
 im\overline{\Psi}(\gamma_5\otimes\sigma_1)\partial_4\Psi, &&
m\overline{\Psi}(\gamma_k\otimes1)\partial_k\Psi,\nonumber\\
 & m\overline{\Psi}(\gamma_k\otimes\sigma_3)\partial_k\Psi, &&
 im\overline{\Psi}(1\otimes1)\partial_4\Psi, && im\overline{\Psi}
(1\otimes\sigma_3)\partial_4\Psi, && im\overline{\Psi}(\gamma_5
\gamma_4\otimes\sigma_2)\partial_4\Psi, \nonumber \\
 & im\overline{\Psi}(\gamma_5\gamma_k\otimes\sigma_2)\partial_k\Psi,
 && m^2\,\overline{\Psi}(1\otimes1)\Psi, && m^2\,\overline{\Psi}(1
 \otimes\sigma_3)\Psi, && im^2 \, \overline{\Psi}(\gamma_4\otimes1)
 \Psi, \nonumber \\
 & im^2\:\overline{\Psi}(\gamma_4\otimes\sigma_3)\Psi, &&
 m^2\:\overline{\Psi}(\gamma_5\otimes\sigma_1)\Psi, && m^2\:\overline{
\Psi}(\gamma_5\gamma_4\otimes\sigma_2)\Psi && \label{eq:dimension4op}
\end{align}
Please note that the lists of operators given here do not contain only
mutually independent operators. Redundancy is potentially present
in these terms, however, in an effort to construct lattice chiral
perturbation theory, having redundant operators poses no major
hurdles \cite{lee1999partial}.


\section{Interacting KW Action\label{sec:interacting_symanzik_action}}

In the interacting case, the only but important difference is the
introduction of gluon fields through covariant derivatives and gauge
action. In terms of gluon link variables $U(x,x+\mu)$ the KW action
becomes,
\begin{equation}
\begin{split}
\mathcal{S} = &\frac{1}{2}\sum\limits^3_{k=1}\sum\limits_x\biggl[
\begin{aligned}[t]
 & \overline{\psi}_x\gamma_kU(x,x+k)\psi_{x+\hat{k}}-\overline{\psi}_{x
+\hat{k}}\gamma_kU^\dag(x,x+k)\psi_{x}\biggr] \\
 & -\frac{i}{2}\sum_x\biggl[\overline{\psi}_x\gamma_4U(x,x+\hat{e}_4)
\psi_{x+\hat{e}_4}+\overline{\psi}_{x+\hat{e}_4}\gamma_4U^\dag(x,x+
\hat{e}_4)\psi_{x}\biggr] \\
 & +m_0\sum_x\overline{\psi}_x\psi_x \\
 & - \frac{i}{2}\sum^3_{k=1}
\sum_x \biggl[\overline{\psi}_x\gamma_4U(x,x+k)\psi_{x+\hat{k}}+
\overline{\psi}_{x+\hat{k}}\gamma_4U^\dag(x,x+k)\psi_{x}-
2\overline{\psi}_x\gamma_4\psi_x\bigg]
\end{aligned}
\label{eq:interacting_lattice_action}
\end{split}
\end{equation}
The realization of point-splitting also gets altered accordingly
with the addition of link variables. To achieve a gauge-invariant
point-split KW action, we define our point-splitting,
\begin{eqnarray}
u_x &=& \frac{1}{2}e^{i\frac{\pi}{2}x_4}\left[\psi_x+\frac{i}{2}(U(x,x_b)
\psi_{x_b}-U(x,x_f)\psi_{x_f})\right],
\label{eq:interacting_point-splitting_up}\\
d_x &=& \frac{1}{2}\Gamma e^{-i\frac{\pi}{2}x_4}\left[\psi_x-\frac{i}{2}
(U(x,x_b)\psi_{x_b}-U(x,x_f)\psi_{x_f}) \right]
\label{eq:interacting_point-splitting_down}
\end{eqnarray}
where $x_b=x-e_4$ and $x_f=x+e_4$. However, Eq.
(\ref{eq:interacting_point-splitting_up}) and Eq.
(\ref{eq:interacting_point-splitting_down}) can be inverted to write
$\psi_x$ in terms of $u_x$ and $d_x$ and the relation connecting
$\psi_x$ and $u_x$, $d_x$ remains unchanged as in Eq.
(\ref{eq:inverse_point-splitting_psi}) and Eq.
(\ref{eq:inverse_point-splitting_psibar}). Defining taste isospinor
$\Psi$ as before in Eq.~(\ref{eq:isospinor}) we get the point-split
interacting KW action in Eq.
(\ref{eq:creutz_point-split_interacting_action}). The order of
appearance of various terms in the interacting action remains the
same as stated in Eq.~(\ref{eq:labeling_eqn}).
\begin{equation}
\begin{split}
\mathcal{S} = & \frac{1}{2}\sum^3_{k=1}\sum_x \biggl[
\begin{aligned}[t]
 & \overline{\Psi}_x\bigg\{(\gamma_k\otimes\sigma_3)+(- 1)^{x_4}(
\Gamma\gamma_k\otimes i\sigma_2) \bigg\}U(x,x+k)\Psi_{x+k} \\
 & - \overline{\Psi}_{x+k}\bigg\{(\gamma_k\otimes\sigma_3)+(- 1)^{x_4}(
\Gamma\gamma_k\otimes i\sigma_2) \bigg\}U^\dag(x,x+k)\Psi_{x}\biggr]
\end{aligned} \\
 & +\frac{1}{2}\sum_x \biggl[
\begin{aligned}[t]
 & \overline{\Psi}_x\bigg\{-(\gamma_4\otimes\sigma_3)-(- 1)^{x_4}(
\Gamma\gamma_4\otimes i\sigma_2) \bigg\}U(x,x+\hat{e}_4)\Psi_{x+e_4} \\
 & + \overline{\Psi}_{x+e_4}\bigg\{(\gamma_4\otimes\sigma_3)-
(- 1)^{x_4}(\Gamma\gamma_4\otimes i\sigma_2) \bigg\}U^\dag(x,x+
\hat{e}_4)\Psi_{x}\biggr]
\end{aligned} \\
 & - \frac{i}{2}\sum^3_{k=1}\sum_x \biggl[
\begin{aligned}[t]
 & \overline{\Psi}_{x}\bigg\{(\gamma_4\otimes1)-(- 1)^{x_4}(\Gamma
\gamma_4\otimes \sigma_1) \bigg\}U(x,x+k)\Psi_{x+k} \\
 & +\overline{\Psi}_{x+k}\bigg\{(\gamma_4\otimes1)-(- 1)^{x_4}(\Gamma
\gamma_4\otimes \sigma_1) \bigg\}U^\dag(x,x+k)\Psi_{x} \\
 & -2\overline{\Psi}_{x}\bigg\{(\gamma_4\otimes1)-(- 1)^{x_4}(\Gamma
\gamma_4\otimes \sigma_1) \bigg\}\Psi_{x}\biggr]
\end{aligned} \\
 & + m_0\sum_x \overline{\Psi}_{x}\bigg\{(1\otimes\sigma_3)+
(- 1)^{x_4}(\Gamma\otimes i\sigma_2) \bigg\}\Psi_{x}
\end{split}
\label{eq:creutz_point-split_interacting_action}
\end{equation}
The next step to construct the interacting Symanzik action is to
identify the symmetries of the interacting lattice action in
Eq. (\ref{eq:creutz_point-split_interacting_action}). For this
we need to know how the link variables transform under the discrete
symmetry transformations. Various discrete symmetries and their
action on the link variables is summarized in Table
\ref{tab:links_transfrm}. 

\begin{table*}[htb]
\caption{\label{tab:links_transfrm} Symmetry transformations and
their actions on gluon fields $A_\mu(z)=A_\mu^B(z)\tau^B$ and
the link variables $U(x,x+\hat{\mu})$ \cite{Lahiri:2022rlg}. Here too,
the first three transformations are linear while Site-reflection is
anti-linear.  Action on spatial and temporal part shown seperately.}
\begin{ruledtabular}
\begin{tabular}{lll}
 Transformation & $A_\mu(z)$ & $U(x,x+\hat{\mu})$ \\ \hline
 Parity (spatial) &$A_j(z)\xrightarrow{P}-A_j(-\mathbf{z},
z_4)$ & $U(n,n+\hat{j})\xrightarrow{P}U^\dag((-\mathbf{n}-\hat{j},
n_4),(-\mathbf{n},n_4))$ \\
 Parity (temporal) & $A_4(z)\xrightarrow{P}A_4(-\mathbf{z},
z_4)$ & $U(n,n+\hat{e}_4)\xrightarrow{P}U((-\mathbf{n},n_4),(-\mathbf{
n},n_4+ 1))$ \\ \hline
 Charge Conjugation & $A_\mu(z)\xrightarrow{C}-A^T_\mu(z)$ &
$U(n,n+\hat{\mu})\xrightarrow{C}U^*(n,n+\hat{\mu})$\\ \hline
 Time Reversal (spatial) & $A_j(z)\xrightarrow{T}A_j(
\mathbf{z},-z_4)$ & $U(n,n+\hat{j})\xrightarrow{T}U((\mathbf{n},
-n_4),(\mathbf{n}+\hat{j},-n_4))$\\
 Time Reversal (temporal) & $A_4(z)\xrightarrow{T}-A_4(
\mathbf{z},-z_4)$&$U(n,n+\hat{e_4})\xrightarrow{T}U^\dag((
\mathbf{n},-n_4- 1),(\mathbf{n},-n_4))$\\ \hline
 Site Reflection (spatial) & $A_j(z)\xrightarrow{SR}-A^{*}_j(
\mathbf{1-z},-z_4)$&$U(n,n+\hat{j})\xrightarrow{SR}U^T((
\mathbf{1-n}-\hat{j},n_4),(\mathbf{1-n},n_4))$\\
 Site Reflection (temporal) & $A_4(z)\xrightarrow{SR}A^{*}_4(
\mathbf{1-z},-z_4)$ & $U(n,n+e_4)\xrightarrow{SR}U^*((\mathbf{1-n},
n_4),(\mathbf{1-n},n_4+ 1))$\\
\end{tabular}
\end{ruledtabular}
\end{table*}

Combining the transformations of the fermionic fields given in Table
\ref{tab:symm_transfrm} with those of the link variables given in
Table \ref{tab:links_transfrm}, we conclude that the symmetries of
the interacting point-split action are identical to that of the
free point-split action in Eq. (\ref{eq:creutz_point-split_free_action}).
Since no dimension-3 operator involves a link variable, the operators
that qualify for the Symanzik action at dimension-3 remain unchanged.
For readability we rewrite those operators,
\begin{align}
i\overline{\Psi}(\gamma_4\otimes1)\Psi, \qquad i\overline{\Psi}(\gamma_4
\otimes\sigma_3)\Psi, \qquad\overline{\Psi}(\gamma_5\otimes\sigma_1)
\Psi \tag{\ref{eq:symanzik_dimension-3_free}}
\end{align}

All dimension-3 operators diverge in the continuum limit as they are
proportional to the inverse lattice spacing i.e. $a^{- 1}$. Hence care
needs to be taken to absorb these divergences. A proper renormalization
scheme for the KW Symanzik action must absorb all the dimension-3
operators by renormalizing appropriate quantities that couple to
operators of higher dimensions. Quantities to be renormalized may
contain wavefunction, fermionic speed of light and gluonic speed of
light \cite{Capitani:2010nn}.

For dimension-4 and dimension-5 operators that are proportional to
$m$ and $m^2$ respectively, the operators that qualify for
the Symanzik action remain unchanged. Since the rotational and
discrete transformations of $\partial\Psi$ and are identical with
those of $D\Psi$, the operators that qualify for the Symanzik action
in dimension-4 and dimension-5 are obtained by replacing
$\partial$ with $D$. Additionally, in dimension-4 we will get two purely
gluonic terms that are invariant under the same set of symmetries.
At the NLO, thus we get the operators,
\begin{align}
 & \overline{\Psi}(\gamma_4\otimes1)D_4\Psi, && \overline{\Psi}(\gamma_4
\otimes\sigma_3)D_4\Psi, && i\overline{\Psi}(\gamma_5\otimes\sigma_1)
D_4\Psi && \overline{\Psi}(\gamma_k\otimes1)D_k\Psi, \nonumber \\
 & \overline{\Psi}(\gamma_k\otimes\sigma_3)D_k\Psi, && m\overline{
\Psi}(1\otimes1)\Psi && m\overline{\Psi}(1\otimes\sigma_3)\Psi,
 && im\overline{\Psi}(\gamma_4\otimes1)\Psi, \nonumber \\
 & im\overline{\Psi}(\gamma_4\otimes\sigma_3)\Psi && m\overline{\Psi}
(\gamma_5\otimes\sigma_1)\Psi, && m\overline{\Psi}(\gamma_5\gamma_4
\otimes\sigma_2)\Psi, && F_{0k}F_{0k}, \quad F_{jk}F_{jk}
\label{eq:dimension4CovOp}
\end{align}
At the NNLO, i.e. dimension-5, the operators entering the Symanzik
action are,
\begin{align}
 & i\overline{\Psi}(\gamma_4\otimes1)D_4D_4\Psi, && i\overline{\Psi}(
\gamma_4\otimes\sigma_3)D_4D_4\Psi, && \overline{\Psi}(\gamma_5\otimes
\sigma_1)D_4D_4\Psi \nonumber\\
 & i\overline{\Psi}(\gamma_4\otimes1)D_kD_k\Psi, && i\overline{\Psi}(
\gamma_4\otimes\sigma_3)D_kD_k\Psi, && \overline{\Psi}(\gamma_5\otimes
\sigma_1)D_kD_k\Psi \nonumber\\
 & i\overline{\Psi}(\gamma_k\otimes1)D_4D_k\Psi, && i\overline{\Psi}(
\gamma_k\otimes\sigma_3)D_4D_k\Psi, && m\overline{\Psi}(\gamma_4
\otimes1)D_4\Psi\nonumber\\
 & m\overline{\Psi}(\gamma_4\otimes\sigma_3)D_4\Psi, && im\overline{
\Psi}(\gamma_5\otimes\sigma_1)D_4\Psi, && m\overline{\Psi}(\gamma_k
\otimes1)D_k\Psi\nonumber\\
 & m\overline{\Psi}(\gamma_k\otimes\sigma_3)D_k\Psi, && im\overline{
\Psi}(1\otimes1)D_4\Psi, && im\overline{\Psi}(1\otimes\sigma_3)D_4
\Psi\nonumber\\
 & im\overline{\Psi}(\gamma_5\gamma_4\otimes\sigma_2)D_4\Psi, &&
im\overline{\Psi}(\gamma_5\gamma_k\otimes\sigma_2)D_k\Psi, &&
m^2\,\overline{\Psi}(1\otimes1)\Psi \nonumber\\
 & m^2\,\overline{\Psi}(1\otimes\sigma_3)\Psi, && im^2\,\overline{
\Psi}(\gamma_4\otimes1)\Psi, && im^2\,\overline{\Psi}(\gamma_4\otimes
\sigma_3)\Psi\nonumber\\
 & m^2\,\overline{\Psi}(\gamma_5\otimes\sigma_1)\Psi, && m^2\,
\overline{\Psi}(\gamma_5\gamma_4\otimes\sigma_2)\Psi &&
\label{eq:dimension5CovOp}
\end{align}
As in the case of free Symanzik action, redundancy in operators
has not been addressed here.


\section{Discussion}

In this paper we addressed the construction of Symanzik effective
theory of Karsten-Wilczek variant of Minimally doubled fermions.
First we considered the free KW lattice action and its doublers
$u_x$ and $d_x$ at the poles of the fermion propagator are separated
using the Creutz-Misumi point-splitting relations. Using these two
doublers or tastes we form a isospin doublet $\Psi$ and then rewrite
the free KW lattice action. This form of the lattice action was
convenient in identifying the taste space symmetries. The invariance
under discrete and taste space symmetries determined the selection
of the continuum operators up to dimension-5 for the Symanzik
effective action. Subsequently, we construct the Symanzik action
for the interacting KW fermions. Once we determine the transformation
properties of the link variables and continuum gauge fields under
discrete transformations, we select the dimension-4 and dimension-5
operators in a way similar to the free case. In the interacting case,
the dimension-3 operators remain unchanged and two purely gluonic
operators appear at dimension-4. The first step in constructing lattice $\chi$PT for MDF is writing down the Symanzik action. Once the Symanzik action is found the chiral Lagrangian can be constructed using the method of spurion analysis \cite{sharpe1998spontaneous,lee1999partial,rupak2002chiral}. Spurion analysis does not require the knowledge of the numerical values of the low-energy constants in the Symanzik effective action. The fixing of the coefficients of the operators in the chiral Lagrangian comes from lattice simulations.

This effort, however does not come without challenges. An important hurdle associated with Creutz point-splitting is the interpretation of tastes in the continuum limit, but it is still possible to progress and construct a lattice chiral perturbation theory with the Creutz point-splitting. The alternative suggestions made, of late, in \cite{Weber:2025kcl} where the tastes have a natural flavor interpretation in the continuum limit, also seem to be an interesting direction to explore.


\begin{acknowledgments}
One of the authors (KS) acknowledges the insightful discussion on
MDF and chiral Lagrangian he had with Stephen Sharpe and Stephan D\"{u}rr
during the LATTICE2024 conference and helpful email communications
with Johannes Weber. Both the authors acknowledge the discussions on the basics of $\chi$PT and Symanzik effective theory with Dipankar Chakrabarti. This project is supported by research fellowship
provided by HBNI (enrollment number PHYS11202205003).
\end{acknowledgments}

\appendix

\section{Transformations of $\overline{\Psi}_x(\gamma_k\otimes
\sigma_i)\partial_k\Psi_x$}

In this appendix, for the purpose of demonstration we choose a
particular dimension-4 operator for the free KW case and show some
of its transformation properties listed in Table
\ref{tab:symanzik_dimension-4a}. From Table \ref{tab:symm_transfrm},
under parity transformation,
\begin{equation}
x\xrightarrow{P}(\mathbf{-x},x_4), \qquad \partial_\mu\xrightarrow{P}(
\mathbf{-\partial_k},\partial_4) \;\;\;\text{and} \;\;\;
\Psi_x\xrightarrow{P}(\gamma_4\otimes\sigma_3)\Psi_{\mathbf{-x},x_4}
\label{eq:dimension4parity}
\end{equation}
Integrating this operator under parity,
\begin{eqnarray}
\int d^4x\, \overline{\Psi}_x(\gamma_k\otimes\sigma_i)\,\partial_k
\Psi_x & \xrightarrow{P} &
   \int d^4x\overline{\Psi}_{\mathbf{-x},x_4}(\gamma_4\otimes
\sigma_3)(\gamma_k\otimes\sigma_i)(\gamma_4\otimes\sigma_3)(
-\partial_k)\Psi_{\mathbf{-x},x_4} \nonumber \\
  & =& \int d^4x\overline{\Psi}_x(\gamma_k\otimes\sigma_3\sigma_i
\sigma_3)\partial_k\Psi_x
\label{eq:dimension4parityintg}
\end{eqnarray}
The only operator which is invariant under parity corresponds
to $i=3$ \textit{i.e.} $\overline{\Psi}_x(\gamma_k\otimes\sigma_i)
\partial_k\Psi_x$. 

Under Site reflection, referring to the fourth line of Table
\ref{tab:symm_transfrm},
\begin{equation}
\Psi_x \, \xrightarrow{SR} \, \left(T\otimes\sigma_3 \right)
\,\overline{\Psi}^T_{\mathbf{1-x},x_4}. \label{eq:dimension4sitereflct}
\end{equation}
Again, integrating this operator under Site Reflection, we get
\begin{align}
 \int d^4x & \;\Psi^T_{\mathbf{1-x},x_4} \left(T\otimes\sigma_3\right)
 \left(\gamma^*_k\otimes\sigma^*_i \right)\left(T\otimes\sigma_3
\right)\left(-\partial_k \right) \,\overline{\Psi}^T_{\mathbf{1-x},
x_4}\nonumber\\
 = & -\int d^4x\Psi^T_{x}\,\left(T\gamma^T_kT\otimes\sigma_3\sigma^T_i
\sigma_3 \right)\,\partial_k\overline{\Psi}^T_{x}\nonumber\\
 = & \;\;\;\int d^4x\Psi^T_{x}\,\left(\gamma^T_k\otimes\sigma_3\sigma^T_i
\sigma_3 \right)\,\partial_k\overline{\Psi}^T_{x}\nonumber\\
 = & -\int d^4x\partial_k\overline{\Psi}^T_{x}\,\left(\gamma_k\otimes
\sigma_3\sigma_i\sigma_3 \right)\,\Psi^T_{x}\nonumber\\
 = & \;\;\;\int d^4x\overline{\Psi}_{x}\,\left(\gamma_k\otimes\sigma_3
\sigma_i \sigma_3 \right)\,\partial_k\Psi_{x}
\label{eq:dimension4sitereflect_intg}
\end{align}
Here too, the only operator that is invariant under site reflection
is for $\sigma_i=\sigma_3$.

For transformation under the generator $(\gamma_5\otimes\sigma_3)$ of
the group $SU(2)_A$, we expand in small parameter $\epsilon$,
\begin{align}
 & \int d^4x\overline{\Psi}_x\, \left[(1\otimes1)+i\epsilon(\gamma_5
\otimes\sigma_3) \right]\,(\gamma_k\otimes\sigma_i) \,
  \left[(1\otimes1)+i\epsilon(\gamma_5\otimes\sigma_3) \right]\,
\partial_k\Psi_x\nonumber\\
 = & \int d^4x\overline{\Psi}_x\, \left[(\gamma_k\otimes\sigma_i)+
i\epsilon \left\{(\gamma_5\otimes\sigma_3), (\gamma_k\otimes
\sigma_i) \right\} \right]\, \partial_k\Psi_x + \ldots\nonumber\\
 = & \int d^4x\overline{\Psi}_x\, \left[(\gamma_k\otimes\sigma_i)+
i\epsilon(\gamma_5\gamma_k\otimes [\sigma_3,\sigma_i] )\right]\,
\partial_k\Psi_x \label{eq:dimension4su2Agen}
\end{align}
The operator $\overline{\Psi}_x(\gamma_k\otimes\sigma_i)\partial_k
\Psi_x$ is invariant under the generator $(\gamma_5\otimes\sigma_3)$
of the group $SU(2)_A$ for $i=3$.

\section{Tiburzi's Point Splitting}

We commented before that the way point-splitting is achieved for
minimally doubled fermions is not unique. Tiburzi
\cite{Tiburzi:2010bm} used a different form of the KW action than
ours given in Eq.~(\ref{eq:free_lattice_action}) and consequently
his point-splitting is different than Creutz-Misumi's
\cite{Creutz:2010qm}. The difference in the action is in the
temporal part of the kinetic term,
\begin{equation}
\begin{split}
 \mathcal{S} =\sum_x \biggl[ \sum^4_{\mu=1} & \bigl[ \:\overline{\psi}
(x) \,\big(\gamma_\mu-i\gamma_4(1-\delta_{\mu4})\big)\,\psi(x+\hat{\mu})
 \\
 &  -\, \overline{\psi}(x+\hat{\mu})\,\big(\gamma_\mu
+i\gamma_4(1- \delta_{\mu4}) \big)\, \psi(x)\bigr] + \overline{\psi}
(x)\left( m_0+3i\gamma_4\right)\, \psi(x)\biggr] \label{eq:A2tibkw}
\end{split}
\end{equation}
The point-splitting suggested by Tiburzi is based on dividing
the Brillouin zone into two parts,
\begin{eqnarray}
 \psi(k)\big|_{k_\mu\in\mathcal{B}} &=& \psi^{(1)}(k) \nonumber \\
 \psi(k)\big|_{k_\mu\notin\mathcal{B}} &=& \gamma_4\gamma_5\psi^{
(2)}(T_{\pi 4}k) \label{eq:A2tibptsplt}
\end{eqnarray}
where $\psi^{(1)}$ and $\psi^{(2)}$ represent the two taste fields
in the momentum space. The shifted momentum is abbreviated as
$T_{\pi 4}k_\mu=\{\mathbf{k}, (k_4+\pi)\mod{2\pi}\}$. Thus the
division of the Brillouin zone is based only on the temporal
component of the momentum. In this manner, two halves of the
Brillouin zone are attributed to the two tastes. Similar to
Eq. (\ref{eq:isospinor}), an isospinor is constructed containing
the two momentum space taste fields,
\begin{equation}
\Psi(k)=\begin{pmatrix} \psi^{(1)}(k) \\ \psi^{(2)}(k) \end{pmatrix}
\label{eq:A2tibiso}
\end{equation}
The Karsten-Wilczek action in the momentum space can be written
using this isospinor as,
\begin{equation}
\mathcal{S} = \int_{\mathcal{B}}\frac{d^4k}{(2\pi)^4}\overline{
\Psi}(k)\bigg[\sum_\mu i(\gamma_\mu\otimes 1)\sin k_\mu 
-i(\gamma_4\otimes\tau_3)\sum_j(\cos k_j- 1)\bigg]\Psi(k)
\end{equation}
where the tensor product is between the spinor space and the taste
space, as before. In the configuration space the field associated
with the two tastes is the infinitely time-smeared,
\begin{eqnarray}
\delta_{\mathcal{B}}\psi_x &=& \int_{\mathcal{B}}\frac{dk_4}{2\pi}
\sum_{y_4}e^{ik_4(x_4-y_4)}\, \psi_{\mathbf{x},y_4}
\label{eq:A2tibBfld} \\
\delta_{\overline{\mathcal{B}}}\psi_x &=& \int_{\mathcal{\overline{B}}}
\frac{dk_4}{2\pi}\sum_{y_4}e^{ik_4(x_4-y_4)}\, \psi_{\mathbf{x},y_4}
\label{eq:A2tibBbarfld}
\end{eqnarray}
Since the taste fields involve summation over all lattice points in
the temporal direction, they are infinitely non-local. When performing
lattice computations, such non-localities may increase the
computational costs and can cause issues with the interpretation
of the hadrons created from such non-local quark fields. Additionally,
problems are expected while extrapolating the lattice computation
results to infinite spacetime. Hence, we choose to use Creutz-Misumi's
point-splitting over Tiburzi's as the non-locality in Creutz
point-splitting is finite.

\bibliography{kw_sym_eft} 
\end{document}